\documentclass[prletter,showpacs,twocolumn,typeset,superscriptaddress]{revtex4}
\usepackage{amsfonts}
\usepackage{amsmath}
\usepackage{amssymb}
\usepackage{graphicx}
\usepackage{epsfig}
\usepackage{citesort}

\usepackage{dcolumn}
\usepackage{bm}

\begin{document}

\preprint{APS}

\title{Concealed Quantum Information}
\author{Luis Roa}
\affiliation{Center for Quantum Optics and Quantum Information,
Departamento de F\'{\i}sica, Universidad de Concepci\'{o}n, Casilla
160-C, Concepci\'{o}n, Chile.}

\date{\today}

\begin{abstract}
We study the teleportation scheme performed by means of a partially
entangled pure state. We found that the information belonging to the
quantum channel can be distributed into both the system of the
transmitter and the system of the receiver. Thus, in order to
complete the teleportation process it is required to perform an
\textit{unambiguous non-orthogonal quantum states discrimination}
and an \textit{extraction of the quantum information} processes.
This general scheme allows one to design a strategy for concealing
the unknown information of the teleported state. Besides, we showed
that the \textit{teleportation} and the \textit{concealing the
quantum information} process, can be probabilistically performed
even though the bipartite quantum channel is maximally entangled.
\end{abstract}

 \maketitle

Quantum teleportation is a protocol which allows transferring
unknown information codified in a pure state, from one quantum
system to another similar one \cite{Bennett}. In order to teleport a
pure state being in a two dimensional Hilbert space, one requires a
Bell state (ebit) and two units with classical information (2 bits).
These resources, an ebit and two bits, complement each other in the
sense that together they allow transferring two unknown real
numbers, codified in a quantum state, between two remote places.
When the quantum channel is a partially entangled pure state, the
process of teleporting an unknown state becomes probabilistic. The
probabilistic teleportation procedure can be implemented by the
transmitter \cite{Roa,Banaszek} or by the receiver \cite{Li,Hsu}.
Throughout this article we will call \textit{channel information}
the probability amplitude of the bipartite entangled state,
$|\tilde{\phi}^+\rangle_{a_2b}$, shared by the transmitter and the
receiver, i.e., $\cos(\theta/2)$ and $\sin(\theta/2)$, see Eq.
(\ref{Bellp}). In the first probabilistic teleportation scheme the
channel information is completely assumed by the transmitter who
must perform an \textit{unambiguous non-orthogonal quantum states
discrimination} (UQSD) process \cite{Peres,Roa2} in order to
complete the process. In the second scheme the channel information
is completely assumed by the receiver who must perform an
\textit{extraction of the quantum information} (EQI) process
\cite{Li} in order to complete the quantum teleportation procedure.

In this article we show that the channel information can be shared
by both the system of the transmitter and by the system of the
receiver. One can realize how this scheme allows designing a
strategy to hide the unknown information codified in the state to be
teleported. Since this process called \textit{concealing the quantum
information} is not unitary and the hidden information is unknown,
it can not be retrieved by means of a unitary process but only by
means of a probabilistic procedure. Besides, we stress the fact
found here that both, the teleportation and the \textit{concealing
the quantum information} process, can be probabilistically performed
even though the bipartite quantum channel is maximally entangled.

First of all, we review the reported probabilistic teleportation
schemes for the particular case of two dimensional Hilbert spaces.
We denote by subindexes $a_1$ and $a_2$ the systems belonging to the
transmitter (laboratory $a$) and by the subindex $b$ the receiving
system. The unknown state, $|\psi\rangle_{a_1}$, to be teleported is
codified on the $a_1$ system. The $a_2$ and $b$ systems are
previously prepared in the partially entangled state
$|\tilde{\phi}^{+}\rangle_{a_2,b}$, see Eq. (\ref{Bellp}).

A probabilistic teleportation scheme was reported in Refs.
\cite{Roa,Banaszek} and it can be succinctly described by the
following identity:
\begin{eqnarray}
|\psi\rangle_{a_1}|\tilde{\phi}^{+}\rangle_{a_2,b}
&=&\frac{1}{2}\left[|\tilde{\psi}^+\rangle_a\sigma_x|\psi\rangle_b-|\tilde{\psi}^{-}\rangle_a\sigma_x\sigma_z|\psi\rangle_b\right.  \nonumber \\
&&+\left.|\tilde{\phi}^+\rangle_a|\psi\rangle_b-|\tilde{\phi}^-\rangle_a\sigma_z|\psi\rangle_b\right],
\label{I1}
\end{eqnarray}
where the four linear independent bipartite states are given by:
\begin{eqnarray}
|\tilde{\psi}^{\pm}\rangle_{a}&=&\sin\frac{\theta}{2}|0\rangle_{a_{1}}|1\rangle_{a_{2}}\pm\cos\frac{\theta}{2}|1\rangle_{a_{1}}|0\rangle_{a_{2}},
\nonumber\\
|\tilde{\phi}^{\pm}\rangle_{a}&=&\cos\frac{\theta}{2}|0\rangle_{a_{1}}|0\rangle_{a_{2}}\pm\sin\frac{\theta}{2}|1\rangle_{a_{1}}|1\rangle_{a_{2}}.
\label{Bellp}
\end{eqnarray}
Without lost of generality we assume $0\leq\theta\leq\pi/2$. We have
written the Pauli operators as
$\sigma_x=|0\rangle\langle1|+|1\rangle\langle0|$,
$\sigma_z=|1\rangle\langle1|-|0\rangle\langle0|$ where evidently
$\{|0\rangle,|1\rangle\}$ is the eigenbasis of $\sigma_z$. So, by
carrying out a UQSD procedure on the (\ref{Bellp}) states, which has
probability of success
\begin{equation}
p_s=1-\cos\theta, \label{p1}
\end{equation}
and transmitting the results (classical information), the receiver
could successfully complete the teleportation process by performing
the appropiate unitary operator. In other words, the process has
probability $p_s$ of teleporting the unknown state $|\psi\rangle$
with fidelity $1$.

Another probabilistic teleportation scheme was first proposed in
Ref. \cite{Li}. It is extended to the $d-$dimensional Hilbert space
case by Li-YI Hsu \cite{Hsu}. That scheme can be understood from the
following expansion:
\begin{eqnarray}
|\psi\rangle_{a_1}|\tilde{\phi}^{+}\rangle_{a_2,b}
&=&\sqrt{p}\frac{|\psi^+\rangle_a\sigma_x|\check{\psi}\rangle_b-|\psi^-\rangle_a\sigma_x\sigma_z|\check{\psi}\rangle_b}{\sqrt{2}}\nonumber\\
&+&\sqrt{1-p}\frac{|\phi^+\rangle_a|\hat{\psi}\rangle_b-|\phi^-\rangle_a\sigma_z|\hat{\psi}\rangle_b}{\sqrt{2}},
\label{I2}
\end{eqnarray}
where $|\phi^\pm\rangle_a$ and $|\psi^\pm\rangle_a$ are the Bell
states, i.e., they are the states of Eq. (\ref{Bellp}) evaluated
with $\theta=\pi/2$. The $p$ probability is read as
\begin{equation}
p=|\langle0|\psi\rangle|^2\sin^2(\theta/2)+|\langle1|\psi\rangle|^2\cos^2(\theta/2).
\end{equation}
Thus, after applying a measurement on the Bell basis of the
$a_1\oplus a_2$ bipartite system and communicating these results to
the receiver, the state of the $b$ system becomes
\begin{equation}
|\check{\psi}\rangle_b=\frac{1}{\sqrt{p}}(\langle
0|\psi\rangle\sin\frac{\theta}{2}|0\rangle_b+\langle
1|\psi\rangle\cos\frac{\theta}{2}|1\rangle_b), \label{s2}
\end{equation}
or
\begin{equation}
|\hat{\psi}\rangle_b=\frac{1}{\sqrt{1-p}}(\langle
0|\psi\rangle\cos\frac{\theta}{2}|0\rangle_b+\langle
1|\psi\rangle\sin\frac{\theta}{2}|1\rangle_b). \label{s1}
\end{equation}
In order to complete the teleportation, the receiver must apply an
EQI process on the (\ref{s1}) or on the (\ref{s2}) state. For
instance, with probability $p$ the outcome state is (\ref{s2}). In
this case the receiver must apply a \textit{Control-U} unitary
operator \cite{Yang}, $\bar{\chi}_{bB}$, with system $b$ being the
control and an auxiliary system $B$, prepared in the $|0\rangle_B$
state, being the target; Namely:
\begin{equation}
\bar{\chi}_{bB}=|0\rangle_{bb}\langle0|\otimes
I_{B}+|1\rangle_{bb}\langle1|\otimes e^{i\sigma_y\arccos(\tan
\frac{{\theta}}{2})},
\end{equation}
with $\sigma_y=-i\sigma_z\sigma_x$, and $I_B$ being the identity of
the Hilbert space of the $B$ system. It is understood that
$e^{i\sigma_y\arccos(\tan \frac{{\theta}}{2})}$ operates on the
Hilbert space of the $B$ system. Since
\begin{small}
\begin{equation}
\tilde{\chi}_{bB}|\check{\psi}\rangle_b|0\rangle_B=\frac{\sin
\frac{\theta}{2}}{\sqrt{p}}|\psi\rangle_b|0\rangle_B
+\frac{\langle1|\psi\rangle\sqrt{\cos\theta}}{\sqrt{p}}|1\rangle_b|1\rangle_B,
\nonumber
\end{equation}
\end{small}
a measurement process performed on the auxiliary system allows
extracting the quantum information, $|\psi\rangle_b$, with
probability $\sin^2(\theta/2)$. Similarly, with probability $1-p$
the outcome is the (\ref{s1}) state; then the receiver must apply,
on the $|\hat{\psi}\rangle_b|0\rangle_B$ state, the transformed
unitary \textit{Control-U} gate $(e^{i\sigma_x\pi/2}\otimes
I_B)\bar{\chi}_{b_1B}(e^{-i\sigma_x\pi/2}\otimes I_B)$, where it is
understood that $e^{\pm i\sigma_x\pi/2}$ takes action on the Hilbert
space of the $b$ system. In this case the probability of
\textit{extracting the quantum information} is also
$\sin^2(\theta/2)$. Thus, the whole success probability in the EQI
procedure is $2\sin^2(\theta/2)$ which is just the probability
(\ref{p1}). Therefore both processes, the teleportation completed by
a UQSD (T-UQSD) process and the teleportation completed by an EQI
(T-EQI) process have the same optimal probability of success.
However, a difference between these schemes is that in the T-EQI
process an extra auxiliary system is required as a physical
resource. Nevertheless it must be emphasized that in the T-UQSD
process the transmitter needs an extra bit for informing the success
or the failure of the measurement process.

Now we show how the teleportation scheme which requires both the
UQSD and the EQI protocols in order to be completed can be used to
design a strategy to conceal unknown quantum information.

The above described processes, T-UQSD and T-EQI, are based on two
different decompositions of the
$|\psi\rangle_{a_1}|\phi^+\rangle_{a_2,b}$ tripartite state, say, on
the four nonorthogonal linear independent states, (\ref{Bellp}), or
on the orthonormal Bell basis. Here we start our analysis by
considering the expansion of the
$|\psi\rangle_{a_1}|\phi^+\rangle_{a_2,b}$ state on the general
nonorthogonal bipartite basis
\begin{eqnarray}
|\tilde{\phi}_{xy}^{\pm}\rangle_{a}&=&\frac{\cos^{x}\frac{\theta}{2}|0\rangle_{a_{1}}
|0\rangle_{a_{2}}\pm\sin^{y}\frac{\theta}{2}|1\rangle_{a_{1}}|1\rangle_{a_{2}}}{\sqrt{\cos^{2x}\frac{\theta}{2}+\sin^{2y}\frac{\theta}{2}}},
\nonumber\\
|\tilde{\psi}_{xy}^{\pm}\rangle_{a}&=&\frac{\sin^{y}\frac{\theta}{2}|0\rangle_{a_{1}}
|1\rangle_{a_{2}}\pm\cos^{x}\frac{\theta}{2}|1\rangle_{a_{1}}|0\rangle_{a_{2}}}{\sqrt{\cos^{2x}\frac{\theta}{2}+\sin^{2y}\frac{\theta}{2}}},
\label{f2}
\end{eqnarray}
where the two real parameters $x$ and $y$ go independently from $0$
to $1$. Thus, we obtain the identity
\begin{widetext}
\begin{equation}
|\psi\rangle_{a_1}|\tilde{\phi}\rangle_{a_2,b} =
\sqrt{p}\frac{|\tilde{\psi}_{xy}^{+}\rangle_{a}\sigma_{x}|\check{\psi}_{xy}\rangle_{b}
-|\tilde{\psi}_{xy}^{-}\rangle_{a}\sigma_{x}\sigma_{z}|\check{\psi}_{xy}\rangle_{b}}{\sqrt{\mu}}
+
\sqrt{1-p}\frac{|\tilde{\phi}_{xy}^{+}\rangle_{a}|\hat{\psi}_{xy}\rangle_{b}-|\tilde{\phi}_{xy}^{-}\rangle_{a}\sigma_{z}|\hat{\psi}_{xy}\rangle_{b}}
{\sqrt{\nu}}, \label{I3}
\end{equation}
\end{widetext}
where the normalization constants are
\begin{eqnarray}
\mu&=&\frac{4p\left(\cos^{2x}\frac{\theta}{2}+\sin^{2y}\frac{\theta}{2}\right)^{-1}}{\left(
|\langle 0|\psi \rangle |^{2}\sin^{2(1-y)}\frac{\theta}{2}+|\langle
1|\psi \rangle |^{2}\cos ^{2(1-x)}\frac{\theta }{2}\right)},
\nonumber\\
\nu&=&\frac{4(1-p)\left(\cos^{2x}\frac{\theta}{2}+\sin^{2y}\frac{\theta}{2}\right)^{-1}}{
\left(|\langle0|\psi\rangle|^{2}\cos^{2(1-x)}\frac{\theta}{2}+|\langle1|\psi\rangle|^{2}\sin^{2(1-y)}\frac{\theta
}{2}\right) }, \nonumber
\end{eqnarray}
and the outcome states are read as
\begin{small}
\begin{eqnarray}
|\check{\psi}_{xy}\rangle _{b} &=&\frac{\langle 0|\psi \rangle \sin
^{1-y} \frac{\theta }{2}|0\rangle _{b}+\langle 1|\psi \rangle \cos
^{1-x}\frac{ \theta }{2}|1\rangle _{b}}{\sqrt{|\langle 0|\psi
\rangle |^{2}\sin ^{2(1-y)} \frac{\theta }{2}+|\langle 1|\psi
\rangle |^{2}\cos ^{2(1-x)}\frac{\theta }{2 }}},
\label{f3}\\
|\hat{\psi}_{xy}\rangle _{b} &=&\frac{\langle 0|\psi \rangle \cos
^{1-x} \frac{\theta }{2}|0\rangle _{b}+\langle 1|\psi \rangle \sin
^{1-y}\frac{ \theta }{2}|1\rangle _{b}}{\sqrt{|\langle 0|\psi
\rangle |^{2}\cos ^{2(1-x)} \frac{\theta }{2}+|\langle 1|\psi
\rangle |^{2}\sin ^{2(1-y)}\frac{\theta }{2 }}}. \label{f4}
\end{eqnarray}
\end{small}

From Eqs. (\ref{I3}), (\ref{f2}), and (\ref{f4}) we can notice that,
if the transmitter defines the parameters $x$ and $y$ to perform an
UQSD process on the $|\tilde{\phi}_{xy}^{\pm}\rangle_{a}$ and
$|\tilde{\psi}_{xy}^{\pm}\rangle_{a}$ states, then the outcome
state, Eq. (\ref{f4}), of the $b$ system gets partial information of
the channel. In this form, the unknown information of the
$|\psi\rangle$ state is concealed by means of the two real
parameters $x$ and $y$. We call this protocol \textit{concealing the
quantum information} (CQI). Since the CQI procedure is probabilistic
and the $|\psi\rangle$ state is unknown, it can only be
probabilistically retrieved, performing an appropriate EQI process
only by the party who knows the $x$ and $y$ parameters. We suppose
that the parameter $\theta$ could be publicly known.

We also notice that in the particular case $x=y=0$ the T-EQI scheme
arises, whereas in the particular case $x=y=1$ the T-UQSD protocol
holds. It is worth emphasizing that, when the quantum channel is
maximally entangled, i.e., $\theta=\pi/2$, the teleportation
procedure is deterministic only in the particular cases $x=y$;
otherwise the teleportation process is probabilistic. This result is
counterintuitive since one would think that, with maximally
entangled states, the process is always deterministic
\cite{Bennett,Roa,Banaszek,Li,Hsu}. Therefore, even in the case of a
maximally entangled channel, the information, $|\psi\rangle$, can be
concealed by choosing two different $x$ and $y$ parameters.

The probability of discriminating conclusively among the four linear
independent non-orthogonal states (\ref{f2}) is
\begin{eqnarray}
p_{\small UQSD}&=&
p(1-|\langle\tilde{\psi}_{xy}^-|\tilde{\psi}_{xy}^+\rangle|)+
(1-p)(1-|\langle\tilde{\phi}_{xy}^-|\tilde{\phi}_{xy}^+\rangle|),
\nonumber\\
&=&
1-\frac{|\cos^{2x}\frac{\theta}{2}-\sin^{2y}\frac{\theta}{2}|}{\cos^{2x}\frac{\theta
}{2}+\sin^{2y}\frac{\theta}{2}}. \label{puqsd}
\end{eqnarray}
This probability, Eq. (\ref{puqsd}), corresponds to the probability
of concealing the quantum information. Its maximum value $1$ happens
for $\cos^x\frac{\theta}{2}=\sin^y\frac{\theta}{2}$, in this case
the (\ref{f2}) states become the Bell states and the outcome states
(\ref{f4}) become the (\ref{s2}) and (\ref{s1}) states. For a given
$\theta$, the minimum value of $p_{\small
UQSD}=2\sin^2(\theta/2)/[1+\sin^2(\theta/2)]$ corresponding to $x=0$
and $y=1$. In other words, $\min\{p_{\small UQSD}\}_{\{x,y\}}$ is
the infimum probability for the process of concealing the quantum
information.

\begin{figure}[t]
\includegraphics[angle=360,width=0.40\textwidth]{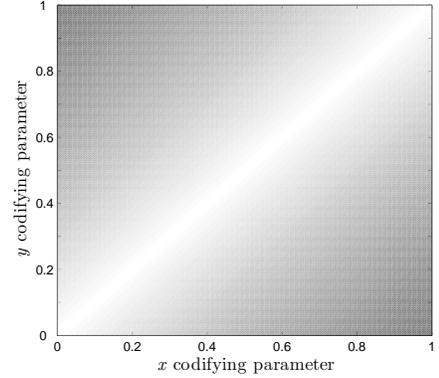}
\caption{Linear black-white degradation of the probability $p_{xy}$
as a function of the $x$ and $y$ secret dimensionless codifying
parameters for; $\theta=\pi/2$  and
$|\psi\rangle=(|0\rangle+\sqrt{2}|1\rangle)/\sqrt{3}$. White color
means probability $1$ whereas the darkest color stands for its
minimum probability value $0.45$.}  \label{figure1}
\end{figure}

As we already see, the probability of concealing the unknown quantum
information, $|\psi\rangle$, by the secret parameters $x$ and $y$ is
equal to the probability of discriminating conclusively among the
linear independent non-orthogonal states (\ref{f2}) and, since the
probabilities of retrieving the quantum information are equal to the
probabilities of extracting the quantum information from the states
(\ref{f3}) and (\ref{f4}), i.e.,
\begin{equation}
\check{p}_{\small EQI} =\frac{\min\{\sin^{2(1-y)}\frac{\theta}{2}
,\cos^{2(1-x)}\frac{\theta}{2}\}}
{|\langle0|\psi\rangle|^{2}\sin^{2(1-y)}\frac{\theta}{2}+|\langle1|\psi\rangle|^{2}\cos^{2(1-x)}\frac{\theta}{2}},
\label{peqi1}
\end{equation}
\begin{equation}
\hat{p}_{\small EQI} =\frac{\min\{\sin^{2(1-y)}\frac{\theta}{2}
,\cos^{2(1-x)}\frac{\theta}{2}\}}
{|\langle0|\psi\rangle|^{2}\cos^{2(1-x)}\frac{\theta}{2}+
|\langle1|\psi\rangle|^{2}\sin^{2(1-y)}\frac{\theta}{2}},
\label{peqi2}
\end{equation}
the whole probability of success, of the process of teleporting
unknown concealed information and thence retrieve it, is given by
\begin{widetext}
\begin{eqnarray}
p_{xy}&=&p(1-|\langle\tilde{\psi}_{xy}^-|\tilde{\psi}_{xy}^+\rangle|)\check{p}_{EQI}
+(1-p)(1-|\langle\tilde{\phi}_{xy}^-|\tilde{\phi}_{xy}^+\rangle|)\hat{p}_{EQI},\nonumber\\
&=& \frac{ 2\min \{\cos ^{2x}\frac{\theta }{2},\sin
^{2y}\frac{\theta }{2}\}\min \{\cos ^{2(1-x)}\frac{\theta }{2},\sin
^{2(1-y)}\frac{\theta }{2}\}}{\cos ^{2x} \frac{\theta }{2}+\sin
^{2y}\frac{\theta }{2}}
\nonumber\\
&&\times\left(\frac{|\langle 0|\psi \rangle |^{2}\cos
^{2}\frac{\theta }{2} +|\langle 1|\psi \rangle |^{2}\sin
^{2}\frac{\theta }{2}}{|\langle 0|\psi \rangle |^{2}\cos
^{2(1-x)}\frac{\theta }{2}+|\langle 1|\psi \rangle |^{2}\sin
^{2(1-y)}\frac{\theta }{2}}+\frac{|\langle 0|\psi \rangle |^{2}\sin
^{2}\frac{\theta }{2}+|\langle 1|\psi \rangle |^{2}\cos ^{2}\frac{
\theta }{2}}{|\langle 0|\psi \rangle |^{2}\sin ^{2(1-y)}\frac{\theta
}{2} +|\langle 1|\psi \rangle |^{2}\cos ^{2(1-x)}\frac{\theta
}{2}}\right).
\label{pt}
\end{eqnarray}
\end{widetext}

We can notice that: the $p_{xy}$ probability depends on the state to
be teleported, when $x=y=0$ or $x=y=1$ the probability $p_{xy}$ is
equal to $p_s$. We can also notice that the T-UQSD and T-EQS
processes can be obtained when $\cos^x(\theta/2)=\sin^y(\theta/2)$
and $\cos^{(1-x)}(\theta/2)=\sin^{(1-y)}(\theta/2)$ respectively.
Figure \ref{figure1} shows a linear black-white degradation of the
probability $p_{xy}$ as a function of the $x$ and $y$ dimensionless
codifying parameters for; $\theta=\pi/2$  and
$|\psi\rangle=(|0\rangle+\sqrt{2}|1\rangle)/\sqrt{3}$. White color
means probability $1$ whereas darkest color stands for the minimum
probability value $0.45$. From Fig. \ref{figure1} clearly we notice
that, in this case of maximally entangled channel, the $p_{xy}$
probability is $1$ on the diagonal $x=y$ only; for other values of
the $(x,y)$ the process is probabilistic and its minimum probability
is different from zero.

In summary, we have studied a general teleportation process which
combines a \textit{conclusive non-orthogonal quantum states
discrimination} process with an \textit{extraction of the quantum
information} scheme. In this form we find a new non-unitary strategy
which allows concealing the quantum information by means of two real
parameters which can be secretly chosen by the party who wants to
protect or hide the information. Since the process of concealing the
unknown quantum information is non-unitary, the process of
retrieving it can not be unitary but it can be probabilistic.
Besides, we showed that both, the teleportation and the
\textit{concealing the quantum information} processes, can be
probabilistically performed even though the bipartite quantum
channel is maximally entangled.

Currently the \textit{deterministic quantum teleportation} scheme
has been experimentally performed with twin-photons and single
photons \cite{Bouwmeester,Boschi,Marcikic,Ursin,Zhang} and with cold
ions $^{40}$Ca$^+$ and $^9$Be$^+$ moving in a linear Paul trap
\cite{Riebe,Barrett} There is also a proposition for teletortating
an atomic state between two cavities using nonlocal microwave fields
\cite{Davidovich}. On the other hand, the \textit{unambiguous
nonorthogonal quantum states discrimination} scheme has been
experimentally demonstrated for two nonorthogonal states of light
\cite{Clarke} and also it has been theoretically proposed with cold
ions in a linear Paul trap \cite{Roa2}. In this way, an experimental
physical implementation of the here proposed \textit{concealing the
quantum information} protocol could consider systems such as single
photons or cold ions.

Further studies of the above described \textit{concealing the
quantum information} scheme can be realized by considering
d-dimensional Hilbert spaces.

\begin{acknowledgments}
This work was supported by Grants: Milenio ICM P02-49F and FONDECyT
N$^{\text{\underline{o}}}$ 1040591.

\end{acknowledgments}

\end{document}